# Implications of Stahl's Theorems to Holomorphic Embedding Pt. 2: Numerical Convergence

Abhinav Dronamraju, *Student Member, IEEE,* Songyan Li, *Student Member,* Qirui Li, *Member*, Yuting Li, *Member*, Daniel Tylavsky, *Life Senior Member,* Di Shi, *Senior Member,* Zhiwei Wang, *Senior Member*

*Abstract*—What has become known as Stahl's Theorem in power-engineering circles has been used to justify a convergence guarantee of the Holomorphic Embedding Method (HEM) as it applies to the power-flow problem. In this, the second part of a two-part paper, we examine implications to numerical convergence of HEM and the numerical properties of a Padé approximant algorithm. We show that even if the convergence domain is identical to the function's domain, numerical convergence of the sequence of Padé approximants computed with finite precision is not guaranteed. We also show that the study of convergence properties of the Padé approximant is the study of the location of the branch-points of the function, which dictate branch-cut topology and capacity and, therefore, convergence rate. We show how poorly chosen embeddings can prevent numerical convergence.

## I. INTRODUCTION

IN the first part of this two-part paper, we showed that convergence of the holomorphic embedding method (HEM) applied to a system of arbitrarily embedded polynomial equations is not theoretically guaranteed, using the power-flow (PF) problem as an exemplar. To guarantee convergence, not only do Assumptions 1.1 in the companion paper have to be satisfied, but the solution point must not lie on the branch cut (BC) with minimum logarithmic capacity, and this BC must not cross the real axis short of the solution point. Even when the necessary conditions are met, Stahl's theorems are silent on the issue of numerical convergence, which is dictated by the convergence properties of the PA, which in turn are affected by the loading pattern of interest and the location of the branch points.

In this paper, we look at the origin of the numerical convergence issues through the lens of the matrix method for calculating the PA [1]. While other PA algorithms will have slightly different numerical characteristics, most, and probably all, of the convergence issues illuminated in this paper will be present and must be understood and respected.

In this paper, using the PF problem as an exemplar, we show that the convergence rate may be easily approximated to practical accuracy, which in turn informs numerical convergence behavior.

While the number of embeddings possible is infinite, we restrict our attention to the two most popular: the classical form and the canonical form, examining these and some variants of the canonical form in the next section. The issues that come to the surface with these embeddings are issues that must be addressed with any embedding. We provide some guidance on embedding decisions that are likely to degrade numerical convergence. We show that the origin of the fundamental numerical limitation of convergence of HEM is that of the PA and that spurious roots are a sign of convergence degradation.

We show that branch-point location affects branch-cut capacity (BCC), which dictates convergence rate. We demonstrate the role that operationally irrelevant branch points can play on the BCC and, hence, on the convergence rate. We show, in essence, that the study of HEM convergence is the study of branch-point location. Finally, we demonstrate that despite the lack of a convergence guarantee in all cases, the robustness of HEM is superior to the continuation power flow in some cases.

## II. HEM VARIATIONS

The four PF problem embeddings discussed in this section will be used to demonstrate how convergence behavior is related to branch-point location and BC topology/capacity. The HEM algorithm is discussed using the classical form embedding as an exemplar, complete with the corresponding recursion relations. Only the defining (embedded) equations are presented for the remaining embeddings.

### A. The Classical Form

The classical form is a scalable form. This means that as the embedding parameter is varied from 0 to 1, the voltage curve generated represents the P-V curve of the system as the PQ load and P generation vary linearly from 0 to full load. With this form, all PV bus voltages are held at their specified values over the entire loading range. For this discussion, VAr limits are ignored, but can and must be accounted for using some bus-type switching algorithm.

#### 1) The Classical Form Embedding

The classical form embedding is defined by (1)-(4),

$$\sum_{k=1}^{N} Y_{ik} V_k(\alpha) = \frac{\alpha S_i^*}{V_i^*(\alpha^*)}, \qquad i \in \{PQ\} \tag{1}$$

$$\sum_{k=1}^{N} Y_{ik} V_k(\alpha) = \frac{\alpha P_i - j Q_i(\alpha)}{V_i^*(\alpha^*)}, \qquad i \in \{PV\} \tag{2}$$

$$V_i(\alpha) * V_i^*(\alpha^*) = |V_i^{sp}|^2 \quad i \in \{PV\} \tag{3}$$

$$V_i(\alpha) = V_{slack}, \quad i \in slack \tag{4}$$

where $V_i(\alpha)$, is the voltage function at bus $i$, $S_i$ is the complex injected power at bus $i$, $Y_{ik}$ are the elements of the admittance matrix, $Q_i$ is the reactive power injection at bus $i$, and $V_{slack}$ is the slack bus voltage. The sets, $\{PQ\}$ and $\{PV\}$, represent the sets of PQ and PV buses, respectively.

*2) Series Representation and Recursion relationship*

The recursion relationship needed to generate the Maclaurin series for the bus voltages and reactive power injected at the PV buses (as functions of the complex-valued embedding parameter, α), will be developed only for the classical formulation. We use the following notation for the voltage and reactive power series.

$$V(\alpha) = V[0] + V[1]\alpha + V[2]\alpha^2 + \cdots + V[n]\alpha^n \quad (5)$$

$$Q(\alpha) = Q[0] + Q[1]\alpha + Q[2]\alpha^2 + \cdots + Q[n]\alpha^n \quad (6)$$

Another phrase for such series is the 'germ' and first term in the voltage germ, $V[0]$, is known as the 'top-of-the germ' or equivalently, 'the reference state.' It is important to remember that in contrast to Stahl's derivation [2], with expansion about infinity, the Maclaurin series is an expansion about the origin.

Substituting (5) and (6) into (1)-(4), and then equating the coefficients of like series terms yields the desired recursion relationships. Only the recursion relationship generated from (1) is shown in (7), where $W_i^*[n-1]$ is the $(n-1)^{th}$ term of the inverse of the voltage series, $V_i^*[k]$. The other recursion relationships are easily generated.

$$\sum_{k=1}^{N} Y_{ik} V_k[n] = S_i^* W_i^*[n-1], i \in \{PQ\} \quad (7)$$

*3) Reference State*

The reference state must be calculated using a different approach than the higher-order terms of the series. The reference state is the solution to (1)-(4) at α=0, namely,

$$\sum_{k=1}^{N} Y_{ik} V_k[0] = 0, \quad i \in \{PQ\} \quad (8)$$

$$\sum_{k=1}^{N} Y_{ik} V_k[0] = \frac{-jQ_i[0]}{V_i^*[0]}, \quad i \in \{PV\} \quad (9)$$

$$V_i[0] * V_i^*[0] = |V_i^{sp}|^2 \quad i \in \{PV\} \quad (10)$$

$$V_i[0] = V_{slack}, \; i \in slack \quad (11)$$

Note that for this classical embedding, solving for the reference state requires solving a no-load power flow, which can be handled in various ways, including using the canonical form, discussed below. Once the reference state is calculated, the recursion relationships are used to find the series coefficients.

*4) Padé Approximants (PAs)*

There is no guarantee that the series generated from the recursion relationships will converge. As discussed in detail in the companion paper, Stahl's theorem [2] states, in short, that under some reasonable assumptions, which experimental evidence suggests the PF equations satisfy (depending on the embedding), the near-diagonal PAs are the maximal analytic continuation of the function described by the series we have obtain. Said another way, the sequence of near-diagonal PAs theoretically converges within the convergence domain. However, there is no guarantee of numerical convergence of any of the numerous PA algorithms in the convergence domain, as observed in [3]. Numerical convergence issues of PAs are discussed in Section III.

*5) Checking for Convergence*

We perform two checks for convergence of the PF algorithm. First, we check for convergence of the sequence of near-diagonal PAs: $|[M-1/M] - [M/M+1]| \le \varepsilon$. Once that convergence test is passed, we check for acceptability of bus power mismatches, since small voltage errors can lead to large mismatch errors at buses with incident branches that have small impedances.

*6) Theoretical Pros and Cons of the Classical Form*

The theoretical advantage of the classical form is that it can be used to analytically represent the P-V curve and therefore can be used to easily estimate the saddle node bifurcation point (SNBP) of the P-V curve [4]. The disadvantage of this form is that to obtain the reference state, one must solve an (arguably more well behaved) no-load PF. The no-load PF solution can be found using any method, such as Newton's method. If HEM is used for calculating the reference state, the canonical form is used.

B. *The Canonical Form*

The canonical form of the HEM embedding is given by (12)-(15),

$$\sum_{k=1}^{N} Y_{ik}^{tr} V_k(\alpha) = \frac{\alpha S_i^*}{V_i^*(\alpha^*)} - \alpha Y_i^{sh} V_i(\alpha), \quad (12)$$
$$i \in \{PQ\}$$

$$\sum_{k=1}^{N} Y_{ik}^{tr} V_k(\alpha) = \frac{\alpha P_i - jQ_i(\alpha)}{V_i^*(\alpha^*)} - \alpha Y_i^{sh} V_i(\alpha), \quad (13)$$
$$i \in \{PV\}$$

$$V_i(\alpha) * V_i^*(\alpha^*) = 1 + \alpha(|V_i^{sp}|^2 - 1), \; i \in \{PV\} \quad (14)$$

$$V_i(\alpha) = 1 + \alpha(V_{slack} - 1), \; i \in slack \quad (15)$$

where $Y_{ik}^{tr}$ are the elements of the transmission admittance matrix, which is simply the admittance matrix with the shunt contribution, $Y_i^{sh}$, removed.

*1) Theoretical Pros and Cons for the Canonical Form*

There are two advantages to the canonical form. The reference state (for systems absent phase shifters) is found by inspection to have all bus voltage values equal to 1+j0 and all PV-bus injected reactive powers equal to zero; thus, no power flow solution is needed for the reference state. Given that shunt elements are often used for power factor correction and voltage support, this formulation also increases shunt element values gradually as the load increases with α, leading to an engineering model that is more reasonable when $\alpha \ne 1$. The disadvantage (in the context of SNBP estimation) is that search methods must be used if the system SNBP is desired since extrapolation of the solution by changing α from 1.0, changes both the voltage set points of the PV and slack buses (in (14) and (15)) and the value of shunt elements in (12)-(13) from their scheduled values.

C. *The Canonical Form with G on the RHS*

This form, proposed in [5], which has been found by the authors to be problematic, will be used to demonstrate a principle about convergence behavior. This form is identical with the canonical form model, but in addition to moving all shunt elements to the RHS, branch conductances are also moved to the RHS. In this form, (12) and (13) are replaced with (16) and (17), respectively,

$$\sum_{k=1}^{N} jB_{ik}^{tr}V_k(\alpha) = \frac{\alpha S_i^*}{V_i^*(\alpha^*)} - \alpha \sum_{k=1}^{N} G_{ik}^{tr}V_k(\alpha) \quad (16)$$
$$- \alpha Y_i^{sh}V_i(\alpha), \quad i \in \{PQ\}$$

$$\sum_{k=1}^{N} jB_{ik}^{tr}V_k(\alpha) = \frac{\alpha P_i - jQ_i(\alpha)}{V_i^*(\alpha^*)} - \alpha \sum_{k=1}^{N} G_{ik}^{tr}V_k(\alpha) \quad (17)$$
$$- \alpha Y_i^{sh}V_i(\alpha), \quad i \in \{PV\}$$

where $B_{ik}^{tr}$ ($G_{ik}^{tr}$) corresponds to the imaginary (real) part of $Y_{ik}^{tr}$.

### D. The Canonical Form with Phase-Shifting Transformers

The advantage of the canonical form is that, at the reference state, all branch flows are zero and the bus voltages can therefore be found by inspection. When phase-shifting transformers are encountered, flows throughout the system may occur at α=0, even though no loads and no shunts exist…and even though all generator buses are controlled to have a voltage magnitude of 1.0 (at no load). Two obvious options are available for dealing with phase-shifting transformers: (a) keep the model of the phase shifters in the admittance matrix on the LHS of the equation and solve a PF problem to obtain the reference state, or (b) move the asymmetrical part of the admittance matrix due to the phase shifters to the RHS, as shown (18) and (19), where $Y_{ik}^{tr\_s}$ ($Y_{ik}^{tr\_as}$) is the symmetrical (asymmetrical) part of the admittance due to the phase shifter model.

$$\sum_{k=1}^{N} Y_{ik}^{tr\_s}V_k(\alpha) = \frac{\alpha S_i^*}{V_i^*(\alpha^*)} - \alpha \sum_{k=1}^{N} Y_{ik}^{tr\_as}V_k(\alpha) \quad (18)$$
$$- \alpha Y_i^{sh}V_i(\alpha), \quad i \in \{PQ\}$$

$$\sum_{k=1}^{N} Y_{ik}^{tr}V_k(\alpha) = \frac{\alpha P_i - jQ_i(\alpha)}{V_i^*(\alpha^*)}$$
$$-\alpha \sum_{k=1}^{N} Y_{ik}^{tr\_as}V_k(\alpha) - \alpha Y_i^{sh}V_i(\alpha), \quad i \in \{PV\} \quad (19)$$

The admittance matrix corresponding to the phase shifter model,

$$Y_{ik} = \begin{bmatrix} A^2y & -Ay\angle\varphi \\ -Ay\angle-\varphi & y \end{bmatrix} \quad (20)$$

is broken down, respectively, as:

$$Y_{ik} = Y_{ik}^{tr\_s} + Y_{ik}^{tr\_as} + Y_{i-k}^{sh}$$
$$= \begin{bmatrix} Ay & -Ay \\ -Ay & Ay \end{bmatrix}$$
$$+ \begin{bmatrix} 0 & Ay - Ay\angle\varphi \\ Ay - Ay\angle-\varphi & 0 \end{bmatrix} \quad (21)$$
$$+ \begin{bmatrix} Ay(A-1) & 0 \\ 0 & y(1-A) \end{bmatrix}$$

where $\varphi$ is the phase-shift angle. The form in (18) and (19) allows the reference state to be obtained by inspection. This form be used to demonstrate a convergence behavior principle.

## III. NUMERICAL CONVERGENCE ISSUES

Assuming theoretical convergence is feasible, the source of numerical convergence issues can be assigned to the following root causes: (1) precision issues with the PA, (2) not matching the embedding form to the problem; (3) branch points added by the embedding form.

### A. Padé Approximant Algorithms

The precise definition of what constitutes a near-diagonal PA was discussed in the companion paper. Based on our observations that the $[M/M + 1]$ PA tends to be the near-diagonal PA with the most reliable performance, we will restrict all analysis and results reported here to $[M/M + 1]$ PAs.

The many methods for calculating PAs [1] can be roughly classified according to their complexity, either $O(M^2)$ or $O(M^3)$. The algorithms that produce solely a numerical value of the function at the desired value of the expansion parameter, $\alpha$, have complexity $O(M^2)$. Those classified as $O(M^3)$, produce a rational approximant. For the Maclaurin series representation of a voltage function of $2M+2$ $c[i]$ coefficients, the $[M/M + 1]$ PA is given by:

$$\sum_{i=0}^{2M+1} c[i]\alpha^i = \frac{a[0] + a[1]\alpha + \cdots a[M]\alpha^M}{b[0] + b[1]\alpha + \cdots b[M+1]\alpha^{M+1}} \quad (22)$$

$$= \frac{(\alpha - a_0)(\alpha - a_2) \ldots (\alpha - a_{2M-2})}{(\alpha - a_1)(\alpha - a_3) \ldots (\alpha - a_{2M+1})} \quad (23)$$

Because we are interested in observing the location of the branch points and the topology of the BCs in this work, we are restricted to $O(M^3)$ algorithms.

#### 1) Numerical Issues with the Padé Approximant Algorithm

In all of the numerical work reported here, we will be using the so-called matrix method, with the following matrix equations for the numerator and denominator coefficients of the PA taken from [1].

$$\begin{bmatrix} c[M] & c[M-1] & \cdots & c[0] \\ c[M+1] & c[M] & \cdots & c[1] \\ \vdots & \vdots & \ddots & \vdots \\ c[2M] & c[2M-1] & \cdots & c[M] \end{bmatrix} \begin{bmatrix} b[1] \\ b[2] \\ \vdots \\ b[M+1] \end{bmatrix}$$
$$= -\begin{bmatrix} c[M+1] \\ c[M+2] \\ \vdots \\ c[2M+1] \end{bmatrix} \quad (24)$$

$$\begin{bmatrix} c[0] & & & \\ c[1] & c[0] & & \\ \vdots & \vdots & \ddots & \\ c[M] & c[M-1] & \cdots & c[0] \end{bmatrix} \begin{bmatrix} 1 \\ b[1] \\ \vdots \\ b[M] \end{bmatrix} = \begin{bmatrix} a[0] \\ a[1] \\ \vdots \\ a[M] \end{bmatrix} \quad (25)$$

Note that because scaling of the numerator and denominator is arbitrary in (22), we arbitrarily set $b[0]=1$. If the matrix in (24) is ill-conditioned, solving (24) can result in significant errors leading to complications (described later). Once (24) is solved, solving (25) requires only a vector matrix multiplication which adds little roundoff error to the result..

The well-known rule of thumb for calculating PAs, regardless of the algorithm used, is to regard as unreliable (approximately) M trailing (guarding) digits of the coefficients of an $[L/M]$ PA. "From a numerical point of view, the Padé approximant derives its capacity to extrapolate certain power series beyond their circle of convergence from using the information contain in the tails of the decimal expansion of the data…." [1] While, for power flow applications, the rule of thumb is usually a very conservative rule, the origin of this rule can be seen through the following example. Consider the function with series expansion shown in (26).

$$f(\alpha) = \sum_{i=0}^{\infty} c[i]\alpha^i \approx \sum_{i=0}^{2M+1=5} \frac{1}{i+1}(-\alpha)^i \qquad (26)$$
$$= 1 - \frac{1}{2}\alpha + \frac{1}{3}\alpha^2 \cdots - \frac{1}{6}\alpha^5$$

Obtaining the coefficients of the denominator polynomial of (22) for a [2/3] PA requires solving:

$$\begin{bmatrix} c[2] & c[1] & c[0] \\ c[3] & c[2] & c[1] \\ c[4] & c[3] & c[2] \end{bmatrix} \begin{bmatrix} b[1] \\ b[2] \\ b[3] \end{bmatrix} = \begin{bmatrix} c[3] \\ c[4] \\ c[5] \end{bmatrix}$$
$$\begin{bmatrix} 3^{-1} & -2^{-1} & 1 \\ -4^{-1} & 3^{-1} & -2^{-1} \\ 5^{-1} & -4^{-1} & 3^{-1} \end{bmatrix} \begin{bmatrix} b[1] \\ b[2] \\ b[3] \end{bmatrix} = \begin{bmatrix} -4^{-1} \\ 5^{-1} \\ -6^{-1} \end{bmatrix} \qquad (27)$$

Solving for $b[3]$ using the determinant method requires evaluating the following determinant,

$$\begin{vmatrix} 3^{-1} & -2^{-1} & -4^{-1} \\ -4^{-1} & 3^{-1} & 5^{-1} \\ 5^{-1} & -4^{-1} & -6^{-1} \end{vmatrix}$$
$$= \frac{1}{3}\left(\frac{1}{4}\frac{1}{5} - \frac{1}{3}\frac{1}{6}\right) + \frac{1}{2}\left(\frac{1}{4}\frac{1}{6} - \frac{1}{5}\frac{1}{5}\right) + \cdots \qquad (28)$$
$$= \frac{1}{3}\left(\frac{1}{20} - \frac{1}{18}\right) + \frac{1}{2}\left(\frac{1}{24} - \frac{1}{25}\right) + \cdots$$
$$= -\frac{1}{3}\left(\frac{1}{180}\right) + \frac{1}{2}\left(\frac{1}{600}\right) + \cdots$$

where only the first two terms of determinant expansion are shown. Notice that the difference inside the parentheses involves two numbers of comparable size; hence the information that needs to be preserved is in the trailing digits of each number. (We would see a similar subtraction of two numbers of comparable size had we performed LU factorization of the matrix in (27).) The number of additional digits needed to represent the difference with the same precision we used for the smallest number in the matrix is on the order of:

$$log_{10}\left(\frac{600}{6}\right) = 2 \qquad (29)$$

Said another way, we must sacrifice 2 guarding digits, reducing the accuracy of the result. Given that round off errors occur, the number of guarding digits must be increased beyond that calculated above, leading to a larger number of guarding digits than this analysis would predict. (The rule of thumb, which would require 3 guarding digits, factors in the practical issues, including roundoff error and matrix conditioning.) Clearly, the number of guarding digits is a function of the matrix numerical content and, to some degree, the algorithm used. (Usually the maximum number of series terms acceptable in PF problems is around 40~50.) While more can be said about the growth of the condition number with increasing matrix size, we have observed that, regardless of the PF problem being solved, the ratio of the series coefficients (see Radius of Convergence (ROC) definition in (30), when that limit exists) tends toward a constant as the number of series terms increases. This leads to ill-conditioning of the matrix, increasing the need for precision.

$$ROC \approx \lim_{i \to large} \left|\frac{c[i]}{c[i+1]}\right| \to constant \qquad (30)$$

The point is this: *even without roundoff error, sooner or later one runs out of precision using finite precision arithmetic, so the number of series terms one can use to approximate any function is limited*. If the problem to be solved is such that convergence is not reached within that limit, then other means must be used to obtain convergence, such as using more precision or restarting the analytic continuation at a point closer to the solution [7].

*2) Spurious Roots*

What happens when the number of guarding digits approaches or exceeds the precision used in the digital implementation of the algorithm? For comparison and simplicity, consider the two-bus model shown in Fig. 1, where the SNBP of the system is 1.7. (The SNBP along the negative real $\alpha$ axis is -1.60.) For a well behaved system, the roots of the [49/50] PA using the classical form embedding are shown in Fig. 2 and Fig. 3 in the $\alpha$ and inverse $\alpha$ planes, respectively. These plots, using extremely high precision, are included as the reference case. (All results presented in this paper are generated using extremely high precision, unless noted otherwise.) Note that because our expansion of the function is about zero, the poles leave the real line in both plots. As we add more terms, the poles gravitate toward the real axis, as observed in the companion paper.

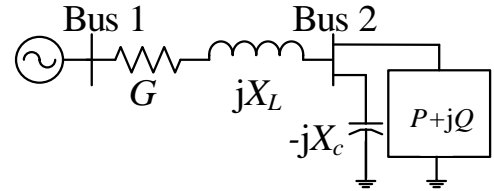

Fig. 1 Two-Bus Model

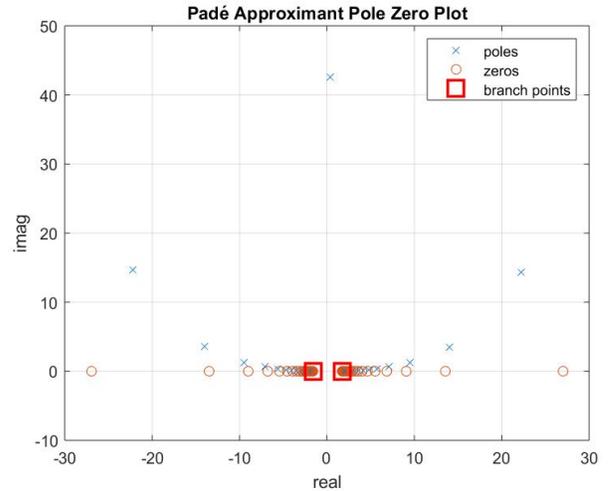

Fig. 2 Two-bus system [49/50] PA roots plot, classical form: $\alpha$ plane (400 digits precision)

If we redo the classical form PF that generated these plots using only double precision, we will be well beyond the precision limitation imposed by the guarding-digit rule and the plot of the roots in the $\alpha$ plane ($\alpha$P) for the [49/50] is shown in Fig. 4 for a much more restricted range. The roots forming (roughly) a circle in Fig. 4 at that ROC of the system (which is approximately the distance from the closest non-spurious pole to the origin, 1.60) are spurious poles caused by lack of precision. Note also that the effect of exceeding the guarding-digit limit on the roots in the inverse $\alpha$ plane (compare Fig. 3 and Fig. 5), can complicate identifying the BC. (Note that only one spurious pole/zero pair, -0.7-j0.0125, is captured in the range of Fig. 5.)

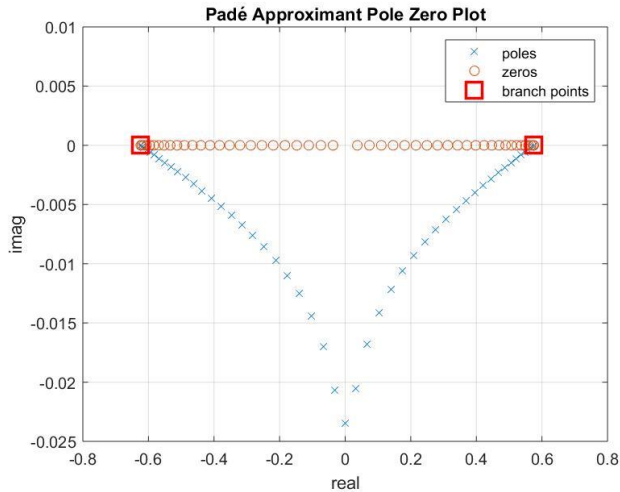

Fig. 3 Two-bus system [49/50] PA roots plot, classical form: inverse $\alpha$ plane (400 digits precision)

If we are not concerned about identifying the BC, do spurious roots help at all? The short answer is that the law of diminishing returns asserts itself, usually sooner rather than later, i.e., these spurious poles do not typically contribute much to the accuracy of the voltage estimate and eventually degrade the accuracy.

Consider the plot of mismatch versus number of series terms of the well-known problematic 43 bus system [8] using the canonical form (adapted from [9]) and shown in Fig. 6. This figure contains a scale on the right showing the number of spurious poles as a function of number of series terms used in building the [$M/M$+1] PA. Observe that once the spurious roots appear, the decrease in bus-power mismatch values versus number of terms ceases and then reverses. Sometimes we do see a small continued improvement once the first spurious poles appear, but that improvement is typically small and short lived.

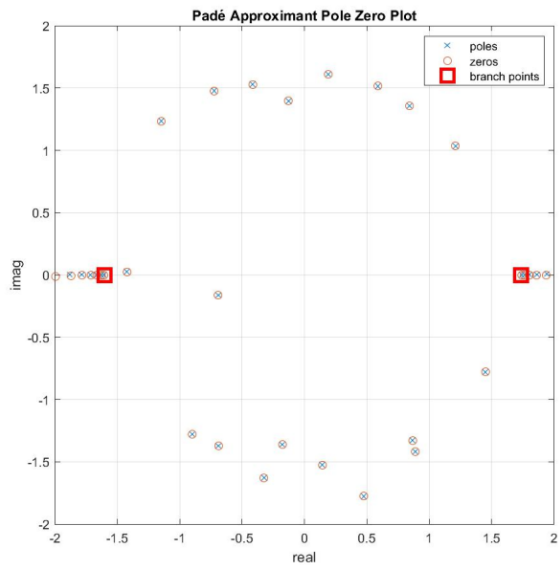

Fig. 4 Two-bus system [49/50] PA roots plot, classical form: $\alpha$ plane (double precision, restricted range)

While the performance of an algorithm executing at double precision is the ultimate litmus test for practical acceptability, the presence of spurious roots can make the PA roots plots appear somewhat chaotic to the untrained eye under near ideal circumstances. Hence, using lower levels of precision will interfere with the ability to draw inferences from the behavior of the subject algorithms, making conclusions difficult and perhaps impossible in some cases. Hence, we use extremely high precision in the remaining simulations, unless otherwise noted.

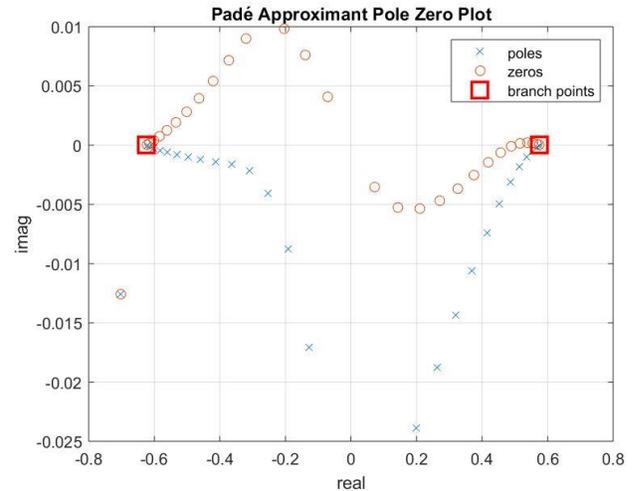

Fig. 5 Two-bus system [49/50] PA roots plot, classical form: inverse $\alpha$ plane (double precision, restricted range)

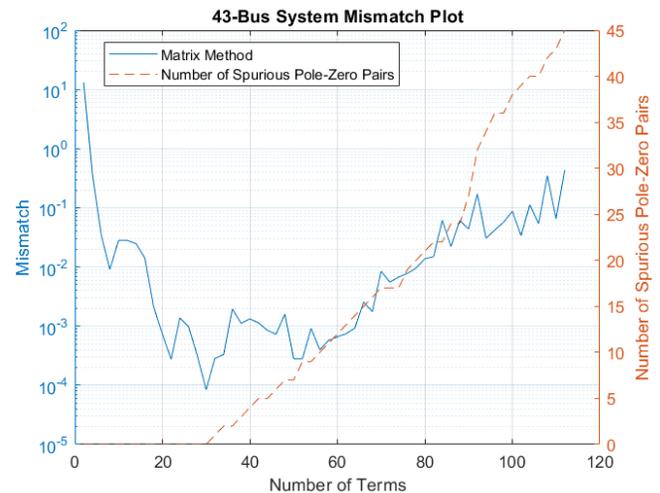

Fig. 6 The 43-bus system mismatch plot at 98% of SNBP (double precision, canonical form, 100 MVA Base)

*3) Reference State Accuracy*

Finally, a note about how the precision of the reference state affects the accuracy of the PA. Unlike Newton's method, where the starting point is an approximation of the solution, the starting point of a HEM algorithm is the solution of the embedded equations at $\alpha = 0$ and the accuracy of the series coefficients is limited by the accuracy of this reference state. If the largest mismatches at the reference state are on the order of 1 MVA, the smallest mismatches one can obtain at the solution point, regardless of the number of series terms and precision used, is on the order of 1 MVA. When the reference state is known by inspection, such as in the canonical form, reference state accuracy is not an issue. However, the classical form requires that we solve a PF problem to find the reference state. Hence the convergence tolerance of the reference state PF must be chosen to be at least as small as the convergence tolerance desired for PF problem itself.

*B. Matching the Objective to the Embedding*

Just as one would use the continuation power flow over, say, the fast-decoupled PF for finding the SNBP, one must be careful to match the HEM form to the problem. If one wants to find the SNBP, the classical form is most efficient. However, if one is solving for the initial state of a transient stability problem, where a significant portion of the real-power load is represented as linear shunt resistance elements, the canonical form is preferred. For such a transient stability model, the reference state PF using the classical form is unlikely to converge because there is likely no solution to the problem. In the reference state PF problem, the only source of real power is the slack bus; hence all power to the linear shunt resistive load elements must flow from the slack bus. For this case, the SNBP is almost certainly less than the reference state loading. Of course, modifying these forms to arrive at hybrid formulations that suit the objective is an option.

As another example, consider the network of Fig. 1 with the following modified values: $V_1=1 \angle 0°$, $P=G=0$, $2X_C=X_L=1$, $Q=-0.1$. Solving this problem with the flat start ($V_2=1 \angle 0°$) using the Newton-Raphson method yields the low voltage solution, as does CPF. While classical HEM easily solves this problem, application of the canonical HEM reveals the problem that confounds the more traditional methods: a pole on the real axis in the αP between origin and the loading point. See the approximate voltage profile produced by canonical HEM in Fig. 7. Just as it is common to move through the BX, XB, and full Newton PF's in search of a solution, having HEM canonical and classical formulations as tools in the toolbelt enhances success.

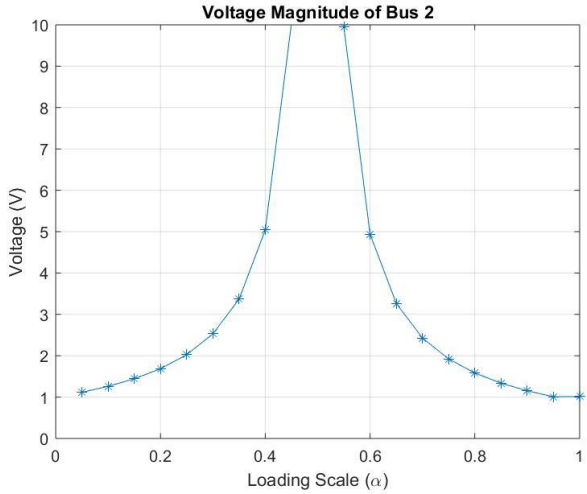

Fig. 7 Modified two-bus system voltage profile $V_1=1 \angle 0°$, $P=G=0$, $2X_C=X_L=1$, $Q=-0.1$ (double precision)

*C. Convergence Behavior Validation*

Stahl's work gives the asymptotic convergence factor (CF) of the PAs cited in the companion paper. The convergence factor for a convergent sequence is defined as $0 \leq |e_{k+1}|/|e_k| < 1$, where $e_k$ is the error metric at iteration $k$ and a value of 1 or greater indicates non-convergence. Asymptotically, the CF for a function expanded about 0, may be derived from [2], and is given in (31). This eq. shows three things: (i) convergence is asymptotically linear; (ii) convergence *rate* (CF$^{-1}$) decreases both with increasing $|\alpha|$, (i.e., as we move away from the point of development) and (iii) increasing branch-cut capacity (BCC).

$$CF(\alpha) = |\alpha|cap(\partial D) + O(\alpha^2) \text{ as } \alpha \to 0 \quad (31)$$

A typical example, which shows the near-linear behavior, is shown in Fig. 8 for loadings at various fractions of the SNBP of the 1354-bus Pegase system. (Note that the SNBP for only the classical embedding represents the SNBP of the power system model; the SNBP for all other embeddings represents the SNBP of that system of equations.)

If we plot the CF of several system models, with $\alpha$ scaled as a fraction of the distance to the SNBP and with BCC ($cap(\partial D)$) as a parameter (see Fig. 9), clearly the increasing value of CF with $|\alpha|$ behavior predicted by (31) is observed. Also, observe that, consistent with (31), the CF is asymptotically linear as $\alpha \to 0$, and is progressively more affected by the $O(\alpha^2)$ terms as $\alpha \to 1$.

Since one of our goals is to experimentally validate (31) and since (31) applies to a single function, we generated Fig. 9 by selecting one bus, finding an accurate value of the voltage at that bus, and then averaging the CF values for magnitude of the bus-voltage error over the sequence of PAs from [0/1] through [24/25] while using 400 digits of precision. We did this for weak and strong buses (identified using modal analysis) and received essentially the same results. We also generated a similar plot using the traditional PF error metric (the maximum value of real or reactive power mismatches at any bus) and the plots had essentially the same behavior.

Observe in Fig. 9 that the slope of each CF curve increases with the *estimated* value of BCC, $cap(\partial D)$, (listed in the legend) as predicted by (31). Ideally, one would use the $cap(\partial D)$ calculated from first principles to validate (31); however, (except for the two-bus case) this approach is beyond the scope of this paper.

An approximate check on the validity of (31) and the results in Fig. 9 for the two-bus case. This case has only two branch points and its BC (of minimal logarithmic capacity) must be a straight-line segment. The BCC of a straight-line segment between $b$ and $a$ is given by [1],

$$BCC(a,b) = |b-a|/4 \quad (32)$$

Observe that the positive and negative SNBP values for the two-bus system corresponding to Fig. 10 are +3.200 and -3.544, respectively, in the αP, or +0.3125 and -0.2822 in the IαP. Observe further that we are scaling the axis in Fig. 9 to be a fraction of the SNBP, which is identical to scaling the α value, call it $\hat{\alpha}$, so that the positive and negative SNBPs for this case are +1 and -0.89696 in the IαP. Hence, for the scaled two-bus problem, the BCC is given by (33), where $\widehat{\partial D}_{2-bus}$ is the BCC for the scaled two-bus problem.

$$cap(\widehat{\partial D}_{2-bus}) = BCC(-0.89696, 1)$$
$$= |1-(-0.89696)|/4 = 0.47474 \quad (33)$$

Using this approach the theoretical convergence factor for the two bus problem at $\hat{\alpha} = 0.01$, is $CF_{2-bus}(\hat{\alpha}=0.01) = 0.00475$, which agrees reasonably well with the experimental value of 0.00451 in the plot of Fig. 9. If we use the magnitude of the bus power mismatch for the two-bus system, the convergence factor is slightly worse (higher) at 0.00484.

Assuming we scale all problems so that the positive SNBP is +1, and neglecting higher order terms in (31), then for sufficiently small $\hat{\alpha}$ values, the CF for any given system, $x$, must be given by (34), and the BCC may be approximated by (35).

$$CF_x(\hat{\alpha}) \approx |\hat{\alpha}| cap(\widehat{\partial D_x}) \tag{34}$$

$$cap(\widehat{\partial D_x}) \approx 100\, CF_x(\hat{\alpha} = 0.01) \tag{35}$$

We used (35) to calculate the BCCs listed in Fig. 9.

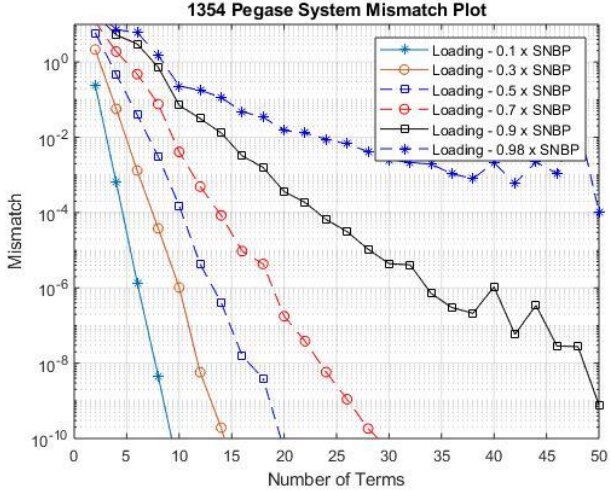

Fig. 8 Mismatch versus number of terms for 1354 Pegase system, classical form (double precision).

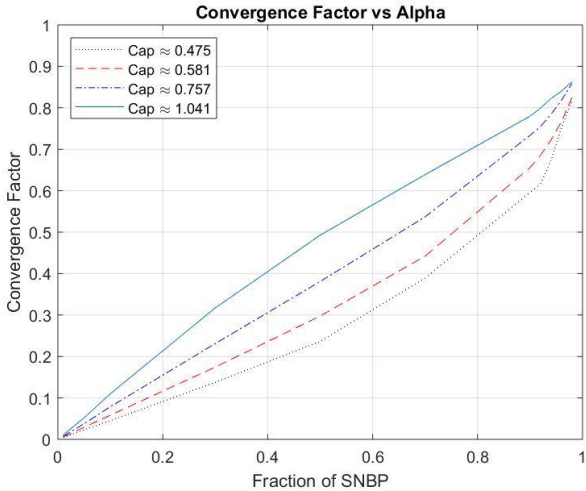

Fig. 9 Convergence factor as function of $\alpha$ and $cap(\partial D)$

The conclusion reached from these experiments is that Stahl's asymptotic convergence factor (adjusted for functions developed around zero) accurately represent the linear convergence behavior of HEM on the PF problem. Given the numerical limitation (that only a limited number of power series terms may contribute to improving voltage-estimate accuracy using PAs), one would expect convergence problems to occur as a higher fraction of SNBP loading is approached (see Fig. 8) and as the BCC increases (see Fig. 9), both of which are observed in practice. Non-convergence for continua approaching and bounded by the SNBP is predicted by these plots, is observed in practice and is clearly a numerical issue that reduces the practical convergence domain (predicted by the theory) to values somewhat below the SNBP. In the next section we examine how the BCC is affected by embedding.

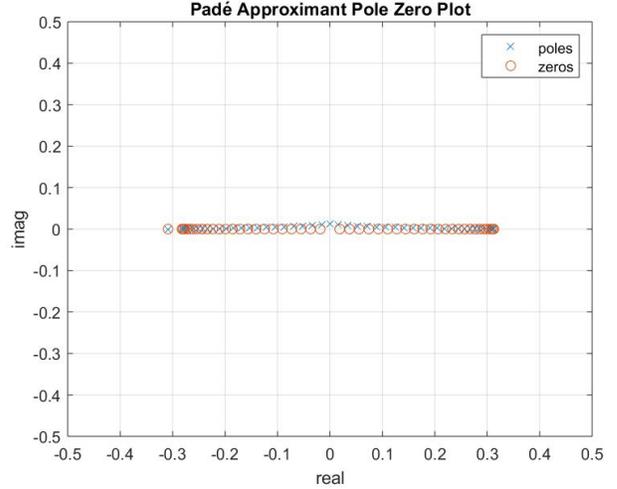

Fig. 10 Branch-cut topology two-bus system, [49/50] PA classical form: inverse-$\alpha$ plane (400 digits of precision)

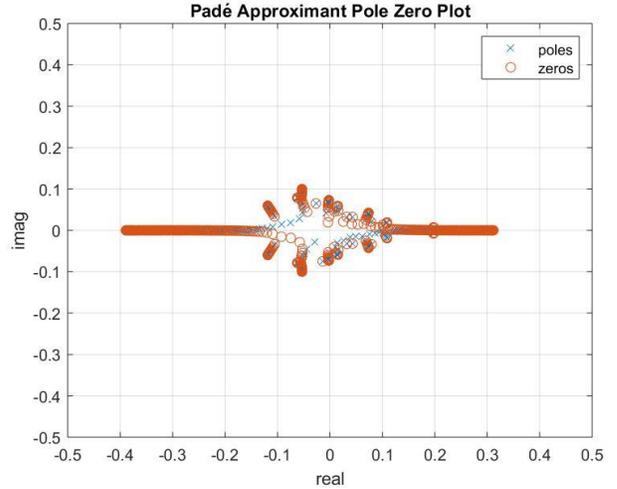

Fig. 11 Branch-cut topology 118-bus system [500/501] PA, canonical form, inverse $\alpha$ plane (1200 digits of precision)

### D. Convergence Behavior as a Function of Embedding

It has been reported that the 'G on the RHS' formulation experiences convergence problems. One of the significant advantages of HEM is the existence of diagnostic tools available to discover the theoretical cause of convergence problems, if not the fundamental modeling error or specific formulation insufficiency. When we move the conductances of the non-shunt branches of the 118-bus system to the RHS according to (16) and (17), the BC approximation of Fig. 11 (with $cap(\widehat{\partial D}_{118}) = 0.581$) becomes that of Fig. 12 (with $cap(\widehat{\partial D}_{118-GonRHS}) = 0.872$), which slows down convergence from 8 terms to 12 terms. While this change is not large, if we change the loading value so that $\hat{\alpha}$ moves closer to 1, and/or change embedding so that the BCC increases, the number of terms needed may exceed the number imposed by the computer precision limitation and convergence lost. *We have observed moving any significant part of the admittance/iteration matrix to the RHS will lead to decreased HEM performance.* Moving shunt elements to the RHS is the notable exception to this rule.

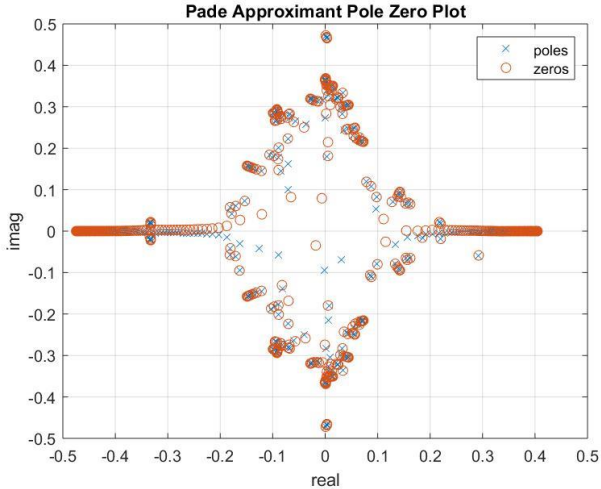

Fig. 12 Branch-cut topology 118-bus system [500/501] PA, canonical form G on RHS, I$\alpha$P (1200 digits of precision)

Observe that, for roots plot in Fig. 12 of the 118-bus 'G on the RHS' formulation, many roots are now accumulating near the branch points close to the imaginary axis, which are now almost as distant from the origin as the SNBP Fig. 12. Because the PA root distribution is equivalent to the equilibrium distribution of electrostatic charge on a 2-D capacitor of the same geometry, the root density is greatest at the branch points most distant from the origin in the I$\alpha$P.

While all roots contribute to the value of the function everywhere, given that roots tend to occur in pole/zero pairs topologically close together, the rate of decrease of their contribution to the value of the function is inversely proportional to their separation. Said another way, a pole and zero closely spaced contribute most to the value near the branch point they are closest to and much less to the function's value at any distant point. Moving portions of the admittance/iteration matrix to the RHS, creates additional (operationally irrelevant) branch points in the complex $\alpha$P, often close to the imaginary axis, and these branch points 'steal roots' from the real valued branch points (SNBP), reducing the accuracy along the real axis. The rate at which the complex-valued branch points 'steal roots' from the operational SNBP increases as these branch points move further from the origin in the I$\alpha$P (move closer to the origin in the $\alpha$P).

Consider the following example of how this can cause non-convergence. We increased only the conductance portions of the impedances of the branches incident on only the weak bus in the 118-bus system, causing that the operating point to move slightly (moving from 31.3% to 31.6% of the system SNBP loading), remaining well within a secure regime. In this case, the closest complex-valued branch point moved closer to the origin in the $\alpha$P, from (about) j2.1 (j 0.47 in I$\alpha$P in Fig. 12) to j0.12 (j 8.6 in the I$\alpha$P). This caused the BCC to change from 0.872 to $cap(\widehat{\partial D}_{118-High\_GonRHS}) = 5.64$, causing the G-on-the-RHS formulation to no longer converge using double precision…though convergence is easily obtained using the canonical form. This demonstrates one possible explanation of why the formulation in (16)-(17) sometimes leads to non-convergence.

The same conclusions, regarding branch point location and moving portions of the iteration matrix to the RHS hold for the canonical form when phase shifters are involved. In fact, the systems of different BCCs in Fig. 9 were created from the 118-bus system, by inserting phase shifting transformers and changing the phase shift angles in the (18)-(19) formulation.

IV. CONCLUSION

We have shown that the number of series terms useful in generating a PA of a voltage function is limited by the finite precision of the computing environment; numerical convergence of HEM in the embeddings studied here is achieved only if the PA can converge to the desire tolerance within this finite number of terms. We provide a convergence factor equation which accurately predicts the convergence rate and a practical means of estimating the branch-cut capacity with sufficient accuracy to be a useful tool. Convergence rate depends on the branch-cut capacity, which in turn depends on the number and placement of branch points, which in turn is dependent upon both the numerical content of the problem and the embedding chosen. We show that moving any significant part of the admittance/iteration matrix to the RHS of the embedding, in general, leads to additional branch points in the complex plane, which can reduce the numerical convergence rate, and ultimately lead to non-convergence.

In our experience, while limitations to HEM convergence exist, the powerful diagnostic tools inherent with HEM have allowed us to easily diagnose the root cause of PF non-convergence when it occurs, remedy the problem, or identify that no solution exists, tasks much more difficult with Newton-based methods. While one objective of this and the companion paper is to show that the putative convergence guarantee often quoted is misleading, HEM converges for many problems where Newton-based methods fail, claiming a rightful niche in the arsenal of PF solvers.

ACKNOWLEDGEMENT

The authors gratefully acknowledge the support for this work provided through the Power System Engineering Research Center (PSERC) by SGCC Science and Technology Program under contract no. 5455HJ160007.

**Abhinav Dronamraju** (Student M'19) was born in Hyderabad, India in 1995. He received the B.Tech. degree in electrical and electronics engineering from Visvesvaraya National Institute of Technology, Nagpur, India. He is currently working toward an PhD degree in electrical engineering at Arizona State University, Tempe, AZ, USA.

**Songyan Li** received his B.E. degree in electrical engineering from North China Electric Power University, Beijing, China in 2013. He received his M.S. degree in electrical engineering from Arizona State University, Tempe, AZ, USA, in 2016. He is now a PhD student at Arizona State University. Songyan Li is a student member of IEEE.

**Qirui Li** received the B.S. degree from Huazhong University of Science and Technology, China, in 2016, and M.S. degree from Arizona State University, US, in 2018, both in electrical engineering. She is currently working at Southwest Electric Power Design Institute Co., Ltd. of China Power Engineering Consulting Group.

**Yuting Li** (S'14–M'16) received her M.S. degree in EE from Arizona State University in 2015. She currently works on the PSS®ODMS product at Siemens PTI. Her research interests include sensitivity-based voltage optimization algorithm, holomorphic embedding power flow algorithm, distribution energy resources (DERs) integration, and FTR trading at electricity market.

**Daniel J. Tylavsky** (LSM) received the B.S. degrees in engineering science and M.S. and Ph.D. degrees in electrical engineering from the Pennsylvania State University, University Park, in 1974, 1978, and 1982, respectively. From 1974 to 1976, he was with Basic Technology, Inc., Pittsburgh, PA, and from 1978 to 1980, he was an Instructor of electrical engineering at Pennsylvania State. In 1982, he joined the Faculty in the School of Electrical, Computer and Energy Engineering, Arizona State University, in Tempe AZ. Dr. Tylavsky is a senior member of IEEE and is an RCA Fellow and NASA Fellow.

**Di Shi** (M'12–SM'17) received his Ph.D. degrees in EE from Arizona State University in 2012. He currently leads the AI & System Analytics Group at GEIRI North America. His research interests include WAMS, AI for power systems, energy storage systems, and renewable integration. He is the Editor of IEEE Transactions on Smart Grid and IEEE Power Engineering Letters, and the recipient of several IEEE PES Best Paper Awards.

**Zhiwei Wang** (SM'18) is the President of GEIRI North America and leads many research activities in the area of AI in power systems, WAMS, and grid modernization.